# Individual structural connectivity defines propagation networks in partial epilepsy

Timothée Proix[1], Fabrice Bartolomei[1,2], Maxime Guye[3,4], Viktor K Jirsa[1]

1 Aix Marseille Université, Inserm, Institut de Neurosciences des Systèmes, UMR_S 1106, 13005, Marseille, France

2 Assistance Publique – Hôpitaux de Marseille, Hôpital de la Timone, Service de Neurophysiologie Clinique, CHU, 13005 Marseille, France

3 Faculté de Médecine de la Timone, centre de Résonance Magnétique et Biologique et Médicale (CRMBM, UMR CNRS-AMU 7339), Medical School of Marseille, Aix-Marseille Université, 13005, Marseille, France

4 Assistance Publique – Hôpitaux de Marseille, Hôpital de la Timone, Pôle d'Imagerie, CHU, 13005, Marseille, France

## Abstract

Neural network oscillations are a fundamental mechanism for cognition, perception and consciousness. Consequently, perturbations of network activity play an important role in the pathophysiology of brain disorders. When structural information from non-invasive brain imaging is merged with mathematical modeling, then generative brain network models constitute in-silico platforms for the exploration of causal mechanisms of brain function and clinical hypothesis testing. We here demonstrate along the example of drug-resistant epilepsy that patient-specific virtual brain models derived from diffusion MRI have sufficient predictive power to improve diagnosis and surgery outcome. In partial epilepsy, seizures originate in a local network, the so-called epileptogenic zone, before recruiting other close or distant brain regions. We create personalized large-scale brain networks for 15 patients and simulate the individual seizure propagation patterns. Model validation is performed against the presurgical stereotactic electroencephalography (SEEG) data and the standard-of-care clinical evaluation. We demonstrate that the individual brain models account for the temporal variability in patient seizure propagation patterns and explain the variability in postsurgical success. Our results show that individual variations in structural connectivity, when linked with mathematical dynamic models, have the capacity to explain changes in spatiotemporal organization of brain dynamics as observed in network-based brain disorders, thus opening up avenues towards discovery of novel clinical interventions.

## Introduction

Neural network oscillations are a fundamental mechanism for the establishment of precise spatiotemporal relationships between neural responses that are in turn relevant for cognition, memory, perception and consciousness. When neurons discharge, the subsequent oscillatory activity propagates through the network recruiting other brain regions, thereby dynamically binding widely distributed sets of neurons into functionally coherent ensembles, hypothesized to represent neural correlates of a cognitive or behavioral content[1]. As the transient synchronization wave evolves, it establishes a spatiotemporal pattern characteristic for cognitive processes[2,3], sensory[4], motor and sensorimotor tasks[5], resting state[6,7] and stimulation paradigms[8]. Alterations of the spatiotemporal organization of these network oscillations play an important role in the pathophysiology of brain disorders. In particular Schizophrenia, Alzheimer and Autism are characterized by reduced functional connectivity and decreased integration of neural processes across the network, whereas Epilepsy and Parkinson's disease show enhanced synchrony of network oscillations responsible for some of the symptomatology. Simple activation paradigms lack the functional complexity to explain the richness of observed spatiotemporal behaviors linked to these brain dynamics[3], leaving it essentially to network processes to explain the origin of the emergent functional and pathological spatiotemporal patterns.

Large-scale brain network models (BNM) emphasize the network character of the brain and merge structural information of individual brains with mathematical modeling. In BNMs, a network region is a neural mass model of neural activity and is connected to other regions via a connectivity matrix representing fiber tracts of the human brain. This form of virtual brain modeling[9,10] exploits the explanatory power of network connectivity imposed as a constraint upon network dynamics and has provided important insights into the mechanisms underlying the emergence of the resting-state networks dynamics[11,12] of healthy subjects, stroke[13] and schizophrenic patients[14]. So far, these studies have exploited generic or averaged connectomes to uncover basic principles of brain network functioning. What yet needs to be demonstrated, however, is the influence of individual structural variations of the Connectome upon the large-scale brain network dynamics of the models. The impact for personalized medicine would be substantial, allowing exploiting the predictive value with regard to the pathophysiology of brain disorders, and their associated abnormal brain imaging patterns. A personalized BNM derived from non-invasive structural imaging data, potentially fit to non-invasive functional imaging data, would allow testing of clinical hypotheses and exploration of novel therapeutic approaches. To explore this capacity of personalized BNM models to serve as a clinical validation and exploration tool in brain network disorders, we here systematically test the virtual brain approach along the example of epilepsy. So far, neural mass models have proven successful in explaining the biophysical and dynamical nature of seizure onsets and offsets[15–19]. Only recently, however, propagation of epileptic seizures started to be studied using BNMs, and was limited to small-scales[20] or absence seizures[21]. Partial seizures have been reported to propagate through large-scale networks in humans[22] and animal models[23]. Around 30 % of the patients with focal epilepsies are drug-resistant. A possible treatment for these patients is the surgical resection of the epileptogenic zone (EZ), a localized region or network where seizures arise, before recruiting secondary networks, called the propagation zone (PZ)[24–26]. As a part of the standard presurgical evaluation, stereotactic EEG (SEEG) are used to help correctly delineating the EZ[27]. Alternative imaging techniques such as structural MRI, M/EEG, and positron emission tomography (PET) help the clinician to outline the EZ. Recently, diffusion MRI (dMRI) and the derived streamlines reflecting the connectivity between different brain regions started being evaluated as well, revealing reduced fractional anisotropy[28,29] and structural alterations in the connectome of epileptic patients[30–32]. However, the usefulness of dMRI in presurgical evaluation of epilepsy remains elusive.

We sharpen our virtual brain hypothesis on the usefulness of personalized large-scale brain network models as clinical tools for the case of epilepsy and will explore their predictive power with regard to the organization of the EZ and the PZ prior to epilepsy surgery. In this article, we build BNMs for a cohort of 15 epileptic patients, systematically simulate the individual seizure propagation patterns and validate the analytical and numerical predictions of the PZ against clinical diagnosis and SEEG signals. Our results demonstrate that personalized virtual brain models reliably predict the PZ for a given EZ. A positive correlation of virtual brain based simulations and surgical outcomes further underlines the causal explanatory power of this approach.

## Results

Partial seizures were recorded with SEEG electrodes in 15 drug-resistant epileptic patients undergoing presurgical evaluation. The clinical characteristics of each patient are given in Supplementary Table 1. Two examples of epileptic seizures of patient CJ are shown in Fig. 1a. On the left of Fig. 1a the seizure is symptomatic and propagates from the EZ, that is, a part of the left lateral occipital cortex (channel highlighted in yellow), to the PZ (channels highlighted in red). On the right of Fig. 1a the seizure is asymptomatic and stays limited to the EZ. Structural and diffusion MRI were preprocessed to construct individualized virtual patients, comprising cortical surface, connectivity matrix, surface and volumetric parcellation, and electrode positions (Fig. 1b-c). We used three different parcellation scales with 70, 140 and 280 cortical regions and 17 subcortical regions, accounting for the uncertainty in the exact size of the EZ due to the sampling issue of SEEG (Fig. 1d).These virtualized patients were then imported into The Virtual Brain[33], an neuroinformatics platform for large-scale brain simulation.

| Patient | Gender | Epilepsy duration (years) | Age at seizure onset (years) | Epilepsy type | Surgical procedure | Surgical outcome | MRI | Histopathology | Side |
|---|---|---|---|---|---|---|---|---|---|
| AC | F | 14 | 8 | Temporo-frontal | Sr | III | Anterior temporal necrosis | Gliosis | R |
| CJ | F | 14 | 9 | Occipital | Sr | III | N | FCD type1 | L |
| CM | M | 35 | 7 | Insular | GK | I | N | NA | L |
| CV | F | 18 | 5 | SMA | Sr | I | N | FCD type2 | L |
| ET | F | 23 | 7 | Parietal | Sr | I | FCD SPC | FCD type2 | L |
| FB | F | 16 | 7 | Premotor | Th | II | N | NA | R |
| FO | M | 45 | 11 | Temporo-frontal | Sr | I | FCD F | FCD type2 | R |
| GC | M | 5 | 28 | Temporal | Sr | III | Temporopolar hypersignal | FCD type1 | R |
| IL | F | 18 | 20 | Occipital | N | NO | N | NA | R |
| JS | M | 11 | 18 | Frontal | Sr | I | Frontal necrosis (post-trauma) | Gliosis | R |
| ML | F | 10 | 17 | Temporal | Gk | II | Hippocampal sclerosis | NA | R |
| PC | M | 15 | 14 | Temporal | N | NO | N | NA | R |
| PG | M | 29 | 7 | Temporal | Sr | I | Cavernoma | Cavernoma | R |
| RB | M | 28 | 35 | Temporal | Sr | III | N | Gliosis | L |
| SF | F | 24 | 4 | Occipital | N | NO | PVH | NA | R |

**Supplementary Table 1: Clinical characteristics of the patients. N, normal; L, left; R, right; Th, thermocoagulation; Gk, Gamma knife; Sr, surgical resection; NO, not operated; PVH, periventricular nodular heterotopia; FCD, focal cortical dysplasia; SPC, superior parietal cortex; F, Frontal; NA, not available.**

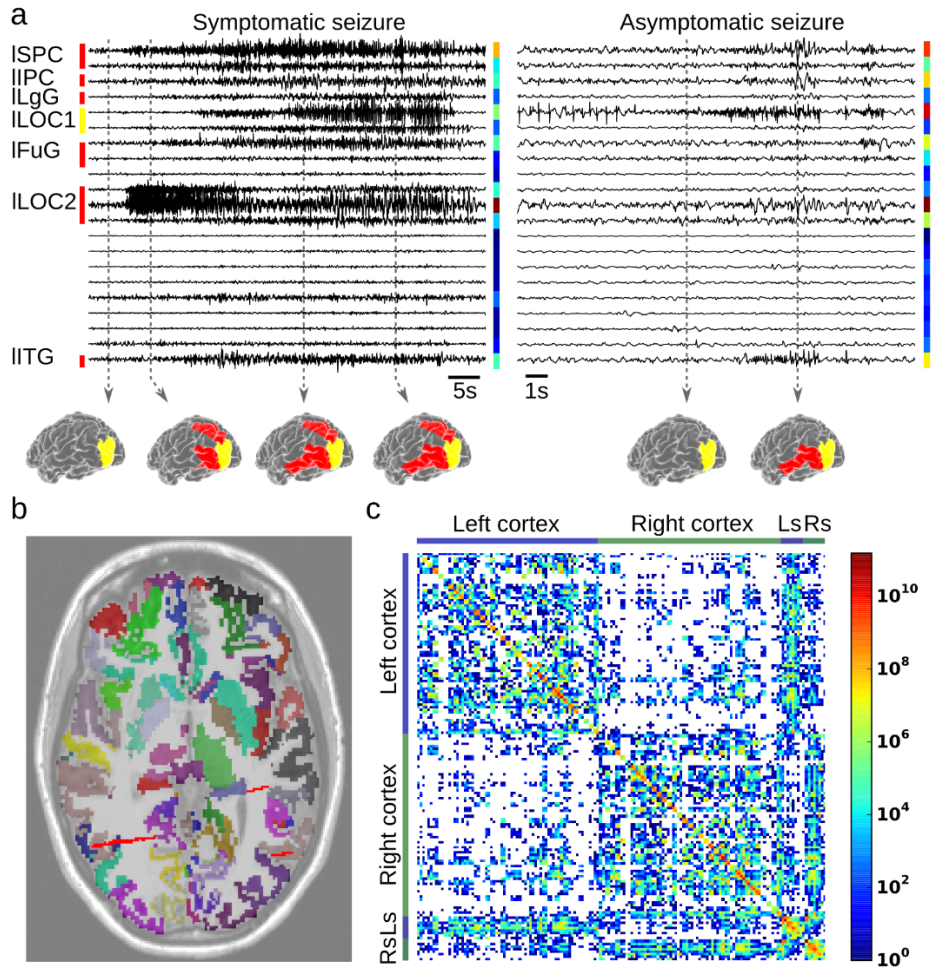

**Figure 1: SEEG data and reconstruction of virtual patient CJ.** (a) Two examples of partial seizures recorded with SEEG in this patient. Left, the seizure propagates from the EZ (yellow) to the PZ (red). Right, the seizure is limited to the EZ. The colorbar shows the power of the signal which is measured. Spatiotemporal activation patterns are shown at different time points of the seizures. lSPC, left superior parietal cortex; lIPC, left inferior parietal cortex; lLgG, left lingual gyrus; lLOC1, part 1 of the left lateral occipital cortex; lFuG, left fusiform gyrus; lLOC2, part 2 of the left lateral occipital cortex; lITG:,left inferior temporal gyrus. (b) Coregistration of the T1 MRI (levels of gray), the parcellation with 157 regions (colors) and the intracranial electrodes (white strips). (c) Connectivity matrix obtained from dMRI for this parcellation.

## Modeling seizure propagation

A BNM[34] was constructed by placing at each node of the parcellation a neural mass model able to reproduce the temporal seizure dynamics and to switch autonomously between interictal and ictal states, the so-called Epileptor model[19]. Epileptors were connected together via a permittivity coupling acting on a slow time-scale, that is on the time-scale of seconds, which is sufficient to describe the recruitment of other brain regions in the seizure[35]. For each patient, the EZ localization was evaluated by a clinical expert and the corresponding regions were set with a high excitability value such that the Epileptors trigger seizures autonomously (Fig. 2a, regions in yellow). Fig. 2b shows an example of a simulation for patient CJ, reproducing both symptomatic and asymptomatic seizures without any parameter changes. For this simulation, the brain regions in the PZ were set with different excitability values to correctly reproduce this recruitment scenario. However, choosing these parameters is a difficult and computationally costly task. In the following, we present a method to estimate the PZ directly from the knowledge of the EZ and the large-scale connectivity matrix.

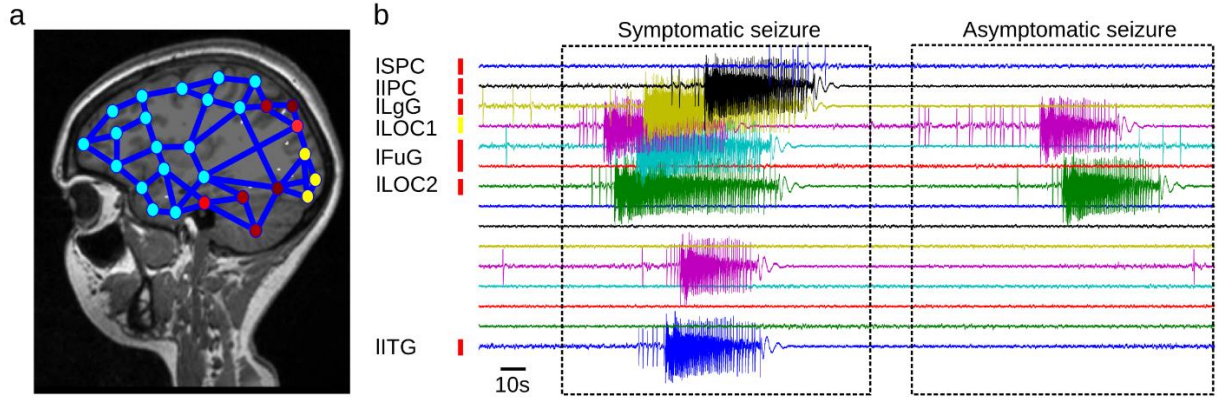

Figure 2: Simulations of the BNM for the patient CJ. (a) A network of Epileptor models is build using the connectivity matrix. The nodes in the EZ are epileptogenic ($\Delta x_{0,i} > 0$, in yellow), the nodes in the PZ have different excitability values ($0.5 > \Delta x_{0,i} > -0.5$, shades of red), while all the other nodes are not epileptogenic (stable state, $\Delta x_{0,i} < 0$, in shades of blue). The blue links represent the anatomical links of the connectivity matrix. (b) Example of time series generated by the simulated BNM with the connectome of patient CJ. Without changing any parameters, the PZ is not always recruited, reproducing the two seizures types of this patient as shown in Fig. 1a.

To predict the PZ as a function of the spatial location of the EZ, the chosen excitability values $\Delta x_{0,i}$, and the connectivity matrix would allow for deeper understanding the conditions leading regions to be recruited in the PZ. The center manifold theorem in non-linear dynamical system theory predicts that the dominating sub-network at the bifurcation is leading in the transition toward the seizure state and can be identified by setting the system at the edge of instability and considering the effect of a small perturbation on the linearized system. We set the regions in the EZ close to the epileptogenic state ($\Delta x_{0,i} = -2.2$, Fig. 3a regions in yellow), and we set all the other regions not epileptogenic with the same excitability ($\Delta x_{0,i} < -2.5$, Fig. 3a, regions in blue). We performed the linear stability analysis on a reduced two-dimensional Epileptor model (see Methods) and computed numerically the corresponding Jacobian matrix. We confirmed the validity of our approach with respect to the simulated Epileptor system (Supplementary Fig. 1a-c).

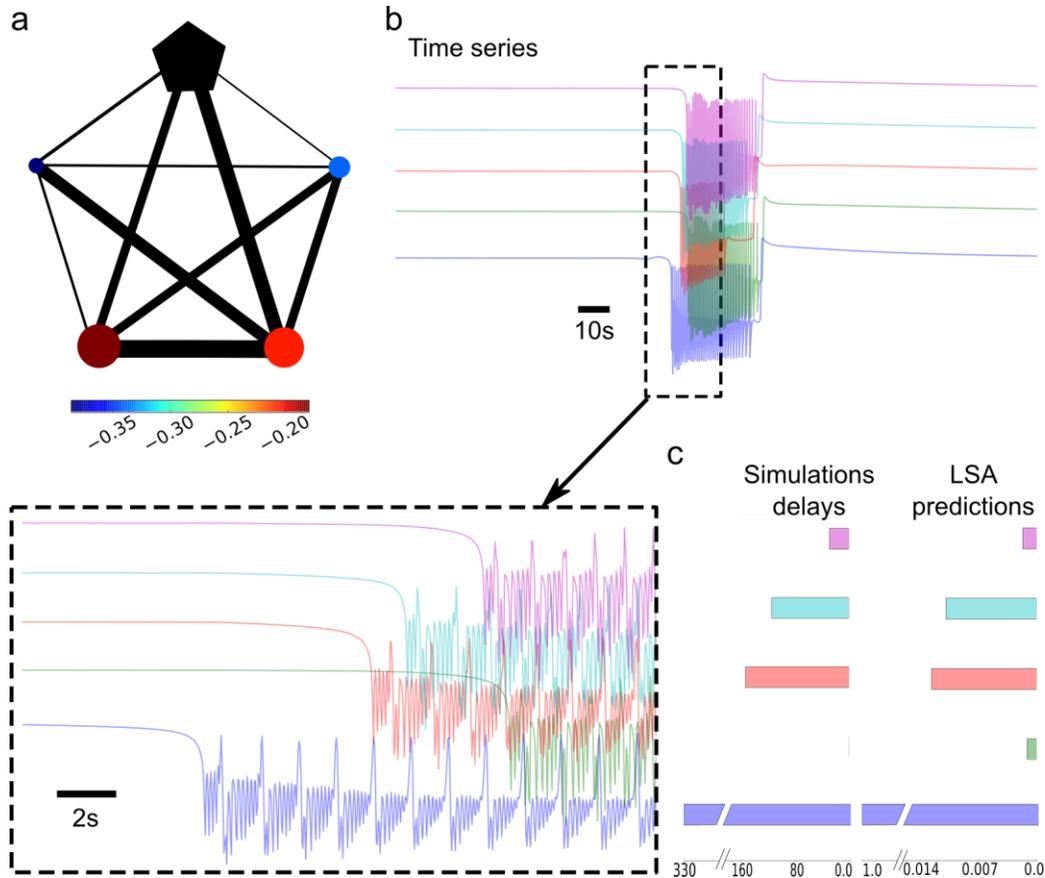

Supplementary Figure 1: Comparison of the prediction of the recruited network by linear stability analysis and by simulations of the whole system. (a) Scheme of a network of five Epileptors. Only the Epileptor represented by a pentagon is epileptogenic ($\Delta x_{0,i} > 0$). The excitability values of the four other Epileptors ($\Delta x_{0,i} < 0$) are represented by the color and the size of the nodes. The strength of the connections are represented by the width of the links. (b) Time series of a simulation for this small network of five Epileptors. Inset is a zoom of the time series, showing the delay before recruitment. (c) Comparison of the delays of recruitment (left) and LSA predictions (right), showing the absolute values of the eigenvector corresponding to the leading eigenvalue as given by the LSA. The eigenvector is normalized such that the first region, here the EZ, has a value of 1. Each Epileptor node has the same color in (b) and (c).

We systematically applied this method for the seizure propagation prediction of the 15 patients. An example of a PZ predicted by the analytical analysis for patient CJ is shown in Fig. 3b. The model prediction of the PZ was compared to (i) the subjective but clinical estimation of the PZ based on all data acquired during the patient evaluation (Fig. 3c), and (ii) the total energy of the SEEG signal over each channel during a seizure to evaluate the PZ[36] (Fig. 3d). For the comparison, two different scores were applied to measure the accuracy of the predicted PZ: (i) a normalized binary score, and (ii) a normalized distance between the predicted PZ values and the reference PZ (i.e., clinical or SEEG estimate). In addition, we used three different parcellations to explore the influence of different sizes of the EZ that is usually unknown. Fig. 3e (patient boxplot) shows the scores obtained for both reference measures and both scores with a parcellation of 158 regions. The average results across all patients and all parcellation types are given in Supplementary Fig. 2 . The individual results are given in Supplementary Table 2 and the abbreviations taxonomy in Supplementary Table 3. In each case, the chance level was found by computing the score for randomly selected regions (see Online Methods), and is shown as a dashed line in Fig. 3 and Supplementary Fig. 2.

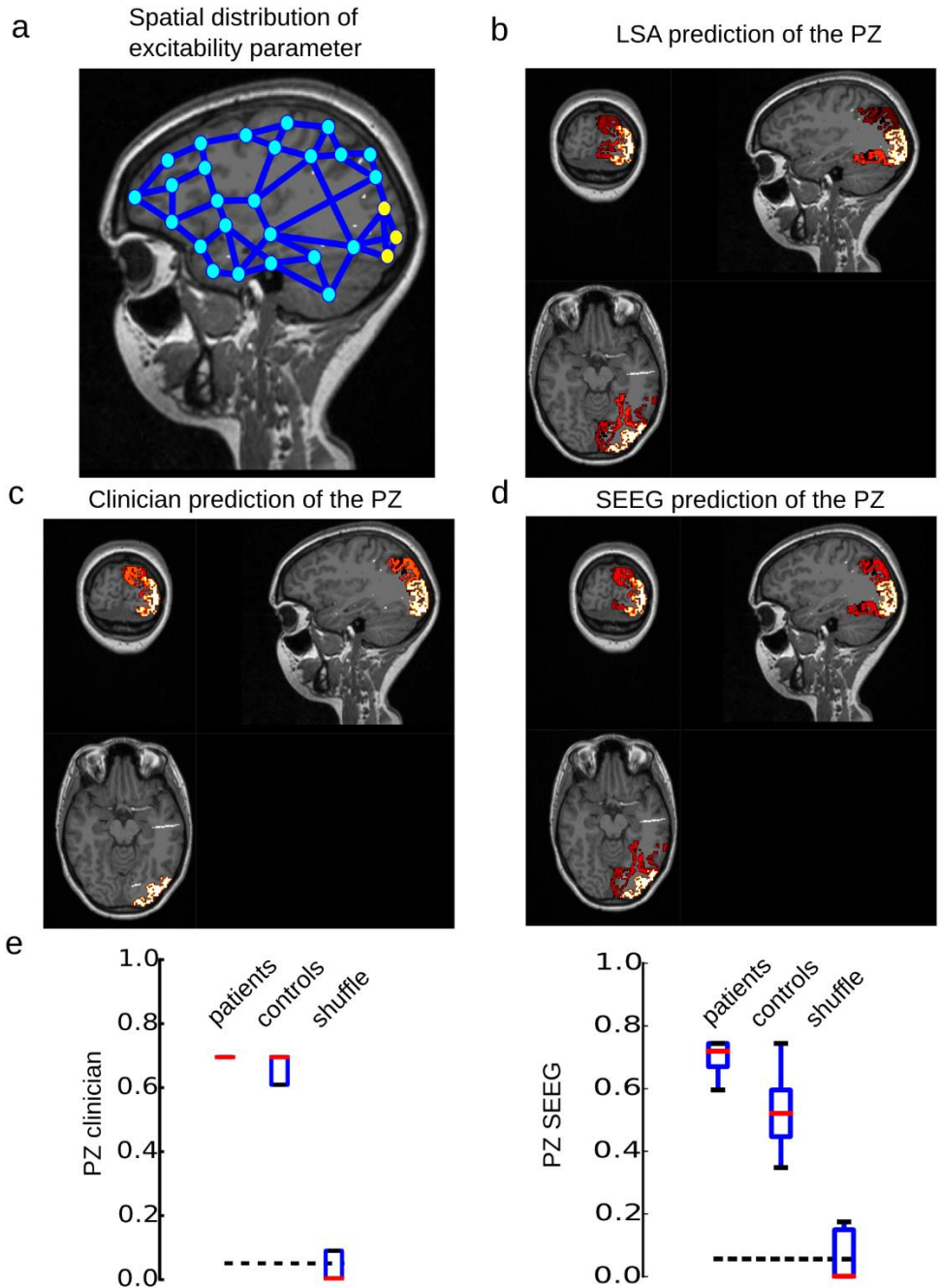

Figure 3: Prediction of the PZ by linear stability analysis. (a) A network of Epileptor models is build using the connectivity matrix. The nodes in the EZ are epileptogenic ($\Delta x_{0,i} < 0$, in yellow), while all the other nodes are equally far from the epileptogenicity threshold ($\Delta x_{0,i} < -0.5$, in blue). Examples of the localization of the EZ (yellow) and the PZ (shade of red) in patient CJ such as found by: (b) LSA using the patient connectome; (c) clinician predictions; (d) SEEG predictions. (e) Results for patient CJ, compared to 5 controls and to shuffled connectivity matrix, for both reference measures (lines) and both scores (columns). The dashed line indicates the level of chance. All boxplots are significantly different (P values <0.01; Mann-Whitney U-test) Left: comparison for patient CJ by region of the number of seizures triggered in a simulation and the analytical prediction. Right: correlation values between simulation and analytical prediction for each patient.

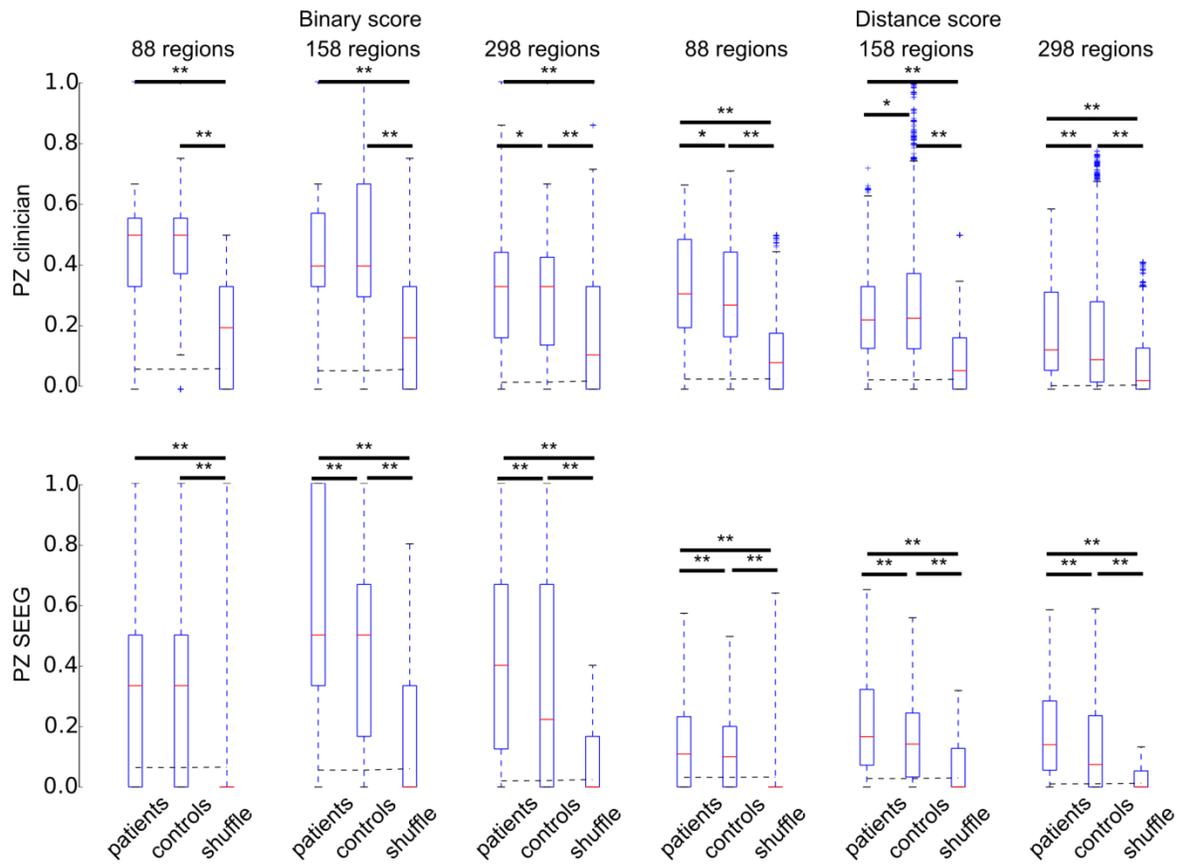

Supplementary Figure 2: Average results over 15 patients, compared to 5 controls and shuffled connectivity matrices of the patients, for both reference measures (lines), both scores (left and right) and different number of regions in the parcellations (each column). The dashed line indicates the level of chance. Significant P values are indicated. ** P <0.01, *P<0.05; Mann-Whitney U-test.

| Patient | EZ location | PZ location | PZ location SEEG | PZ prediction |
| --- | --- | --- | --- | --- |
| AC | rLOFC, rTmP | rSFG, rRMFG, lLOFC | rRMFG, lRMFG | rRMFG, rMOFC, rPOr, rIns, rPut, rPT |
| CJ | lLOC | lIPC, lSPC | lFuG, lPC, lSPC | lFuG, lSPC, lITG, lIPC, lPC, lLgG |
| CM | lIns | lPu, lPoG | lPoG | lPu, lLOFG, lSMG, lPrG, lPOp, lPoG |
| CV | lPCG, lCMFG, lSFG | lPrG, lPoG | lPrG, lSPC, lPoG | lRMFG, lPrG, rSFG, lCACC, lPaC |
| ET | lPCG, lPCunC | lPoG, lSPC | lPoG, lIPC | lICC, lIPC, lSPC |
| FB | rPrG | rPoG, rSP | rCMFG | rPoG, rSF, rCMFG, Pop |
| FO | rAmg, rTP, rLOFC | rHi, rEntG, rMOFC, rMTG, rPHiG; rPOr, rPT, rRMFG, rIns | rFuG, rLOFC, lPHiG, rITG | rIns, rPu, rPOr, rMOFC; rMFG, rHi |
| GC | rAmg, rHi | rSTG, rTmP, rITG, rMOFC, rLOFC | rITG, rTmP | rPHiG, rEntC, rTmP, rFuG, rPal, rTh |
| IL | rLgG, rPHiG | rLOCC, rFuG, rIPC | rHip, rFuG, rIPC, rLOCC, rSPC, rITG | rFuG, rHip, rPC, rLOCC |
| JS | rMOFC, rFP, rRMFG, rPOr | rLOFC, rRMFC, rSFC, rCMFG | rPop, rMTG, rPOr, rLOFC | rCACC, rSFC, rPT, rPrG, rCd, rPop |
| ML | rHi, rAmg | rTh, rCd, rPu, rIns, rEntC, rTmP | rLOFC, rMTG | rPHiG, rTh, rPal, rEntC, rTmP |
| PC | rHi, rFuG, rEntC, rTmP | lFuG, rIPC, rITG, rLOC, rPHiG, rPCunC, rSMG | lFuG, rITG | rLOC, rITG, rLgG, rPHiG |
| PG | rFuG | rEntC, rHi, rAmg, rITG, rMTG | rEntC, rIPC, rHi | rITG, rLOC, rTmP |
| RB | lAmg, lHi, lEntC, lFuG, lTmP, rEntC | lSTG, lMTG, lITG, lIns, lPHiG | lMTG, rMTG, lIns | lPHiG, lITG, lLOC, rHi |
| SF | rLgG, rLOC, rCun, rPC | rPHiG, rSPC, rFuG | rPCunC, lCun, rPHiG | rPC, rFuG, rIPC, rSPC, rPCunC |

Supplementary Table 2: Results of Propagation zone prediction for each patient. Abbreviations are given in Supplementary Table 3.

| | | | |
|---|---|---|---|
| **Amg** | Amygdala | Pal | Pallidum |
| **CACC** | Caudal anterior cingulate cortex | PC | Pericalcarine |
| **Cd** | Caudate nucleus | PCG | Posterior cingulate gyrus |
| **CMFG** | Caudal middle frontal gyrus | PCunC | Precuneus cortex |
| **Cun** | Cuneus | PHiG | Parahippocampal gyrus |
| **EntC** | Entorhinal cortex | PoG | Postcentral gyrus |
| **FP** | Frontal pole | Pop | Pars opercularis |
| **FuG** | Fusiform gyrus | POr | Pars orbitalis |
| **Hi** | Hippocampus | PrG | Precentral Gyrus |
| **ICC** | Isthmus-cingulate cortex | PT | Pars triangularis |
| **Ins** | Insula | Pu | Putamen |
| **IPC** | Inferior parietal cortex | RACC | Rostral anterior cingulate cortex |
| **ITG** | Inferior temporal gyrus | RMFG | Rostral middle frontal gyrus |
| **LgG** | Lingual gyrus | SFG | Superior frontal gyrus |
| **LOCC** | Lateral occipital cortex | SMG | Supramarginal gyrus |
| **LOFC** | Lateral orbito frontal cortex | SPC | Superior parietal cortex |
| **MOFC** | Medial orbito frontal cortex | STG | Superior temporal gyrus |
| **MTG** | Middle temporal gyrus | Th | Thalamus |
| **PaC** | Paracentral cortex | TmP | Temporal pole |

Supplementary Table 3: Abbreviations of cortical and subcortical regions.

To compare our results with the surgical outcome, for each patient, we estimated the number of regions, which were found in the PZ by our model and not explored by SEEG. These are the regions that were not taken into account by the clinicians in the presurgical evaluation (region in green in Fig. 3d left). We found that a large extent of this PZ was correlated with poor seizure prognosis according to the Engel classification[37], classifying postoperative outcomes for epilepsy surgery (Fig. 3d right using clinical prediction, 158 regions and the distance score). A linear fit demonstrated an increasing slope of the linear fit for different parcellation sizes, PZ measures, and scores, but not always significantly different from 0.

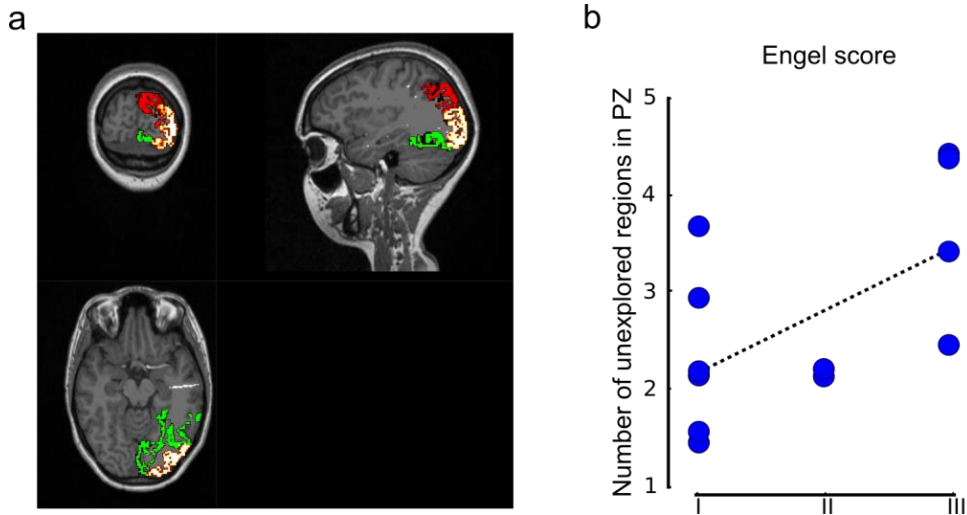

Figure 4: Prediction of the surgical outcome. (a) Example for patient CJ of regions (in green) found in the PZ by the linear stability analysis but not considered in the PZ by the clinical expertise. (b) Comparison for all patients of size of the unexplored regions predicted as in the PZ by the analytical model and the Engel classification (clinical prediction, 158 regions, distance score). The slope of the linear regression is significantly different from 0 (t-test, P value<0.05).

## Surrogates connectivities

To gain more confidences in our results we computed the following surrogate connectivities: We compared each patient score with the average score obtained from five different control connectivity matrices (Fig. 3e and Supplementary Fig. 2, control boxplot, and Supplementary Fig. 3a). The score obtained for the individualized connectivity matrix was slightly better than the control connectivity, but not with significant difference for all subjects. We examined the effects of shuffling the weights of the connectivity matrix, that is, rewiring differently the connectome by changing the topology of the network. We only shuffled the cortical connections inside each hemisphere. The results were always significantly better with the individualized connectivity matrices (Fig. 3e and Supplementary Fig. 2, shuffle boxplot, and Supplementary Fig. 3b). Other connectivities (random Erdös-Renyi networks, Strogatz-Watts small-world networks[38] performed very poorly (result not shown)) compared to the patient connectivity matrices. We also examined the effect of randomly changing up to 20% and 40% of the values of the weights with respect to the original values, therefore respecting the topology of the network (Supplementary Fig. 4, 20% and 40% boxplots). As expected the results did not change significantly. The weights show a log-normal distribution (Supplementary Fig. 5). Other studies have used a normal distribution of weights[39,40], but our results were degraded with a normal distribution of weights (by taking the log of the weight matrices) (Supplementary Fig. 4, log boxplot).

These results suggest that the topology of the connectivity matrix is significantly important to predict the recruited network.

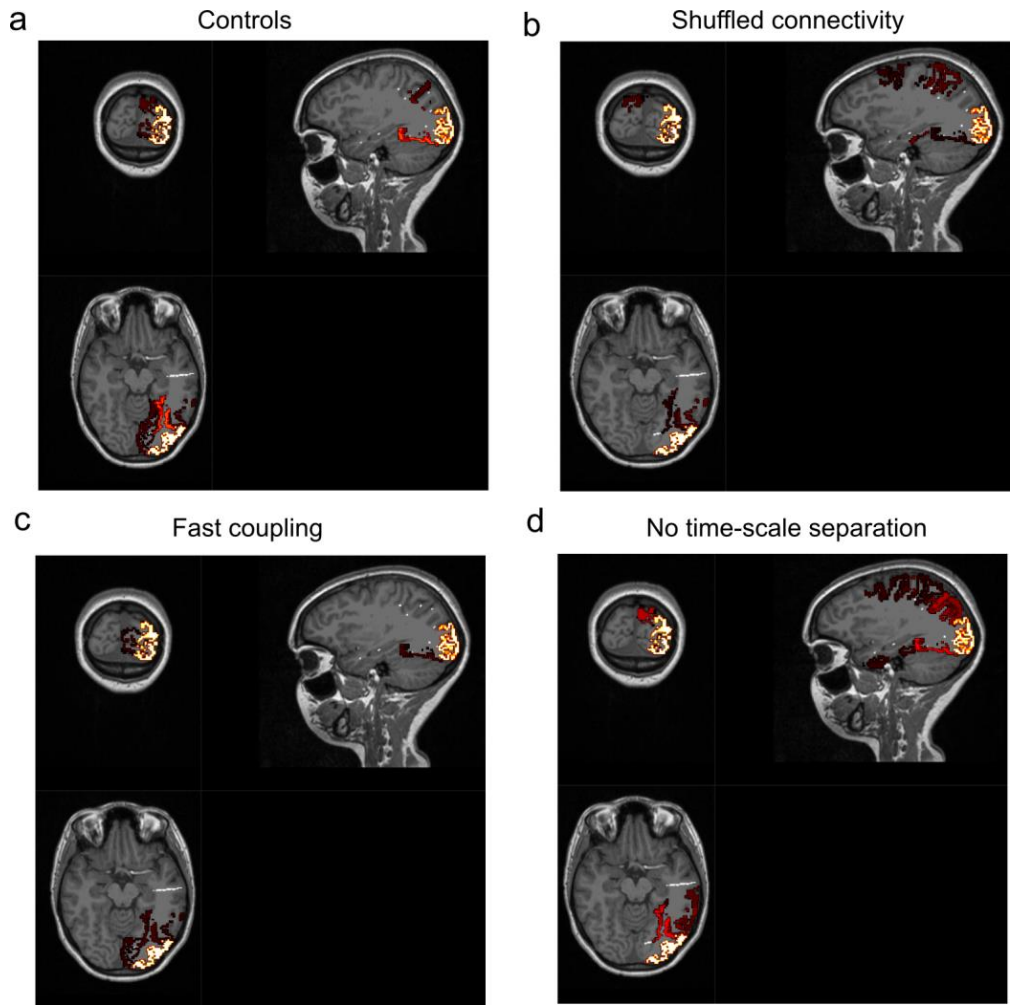

**Supplementary Figure 3:** Examples of using different surrogate connectivities and models for patient CJ. (a) Control connectivity. (b) Shuffled connectivity. (c) Coupling on a fast time scale. (d) No temporal scale difference in the Epileptor model.

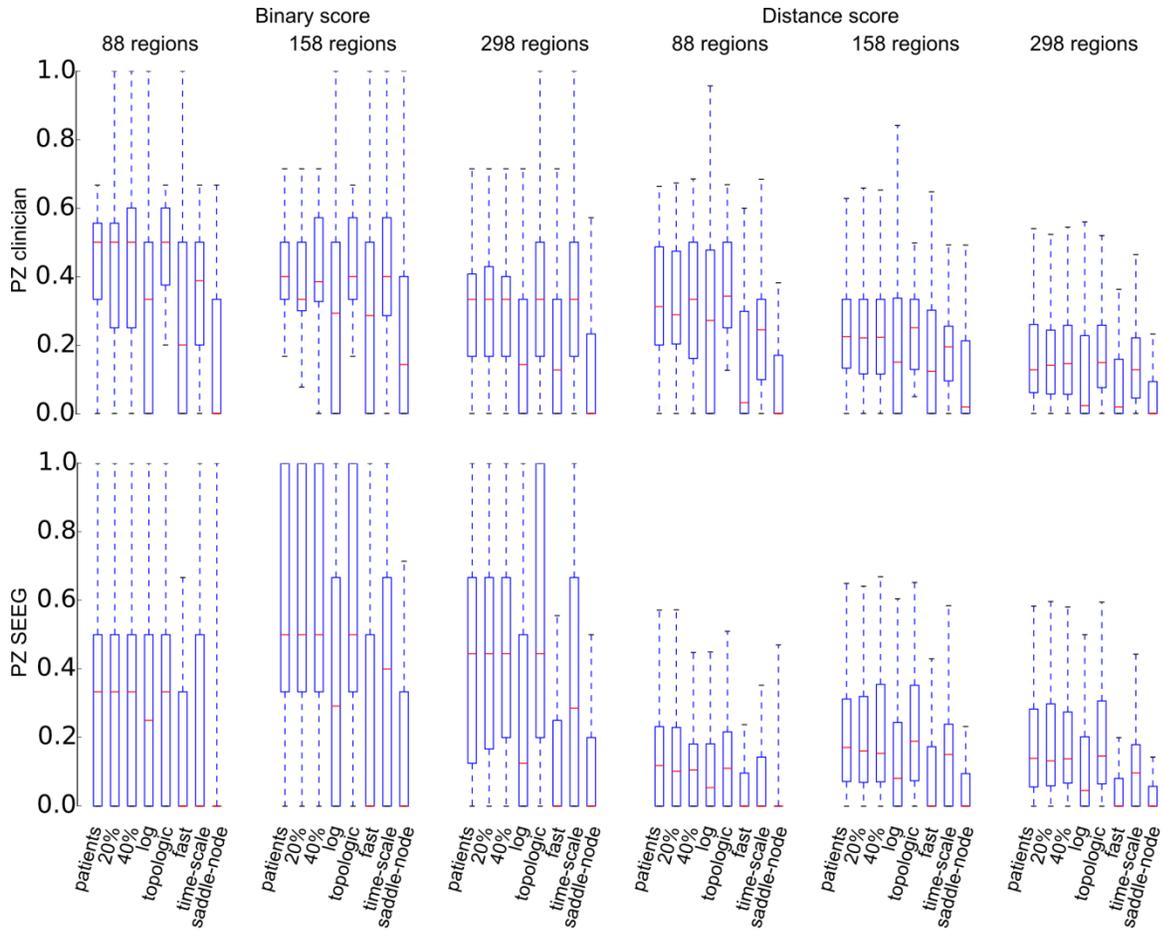

**Supplementary Figure 4:** Score over all subjects for different surrogate connectivities and surrogates models. 20%, changing randomly 20% of the weight values; 40% changing randomly 40% of the weight values; log, logarithm of the weight values; topologic, sum of in and out strength of the EZ; Fast, coupling on a fast time scale; Time-scale, no time scale separation, i.e. $\tau = 1$ in ; Saddle-node, generic saddle-node bifurcation normal form.

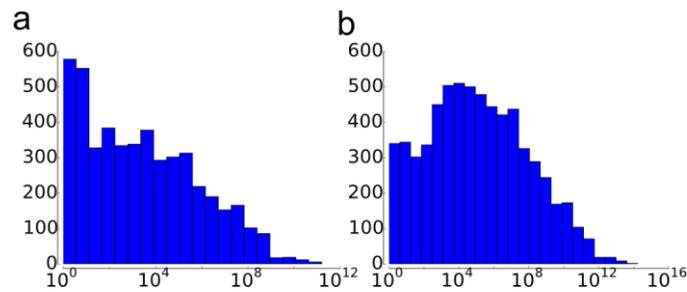

**Supplementary Figure 5:** Distribution of the weights of the connectivity matrix. Note the logarithmic scale on the horizontal axis. (a) For a single patient. (b) Average over the 15 patients

## Surrogate models

We also tested alternative models and coupling functions. First we show that recruited network, as predicted by the BNM, is well approximated by a function of the excitabilities and out-strength of the EZ. The reduced Epileptor is a slow-fast system that we further reduced to a one-dimensional system by projecting the dynamics on the slow manifold. Since the difference coupling leads to small terms at the first order approximation, the fixed point solution of the system does not depend of the coupling function. Furthermore, the linear stability analysis can be simplified by assuming weak couplings (which is the case when normalizing the connectivity matrix to its maximum weight), and is shown to be only a function of the excitabilities and out-strength of the EZ (see Online Methods). In the case of the linear stability analysis (Fig. 3), since excitability of all the nodes outside the EZ is equal, the out-strength directly determines the PZ. Our analytical result is confirmed for a generic connectivity matrix (Supplementary Fig. 1d) as well as for patient connectivity matrices (Supplementary Fig. 4, topology boxplot).

We checked the importance of our assumptions, that is, time-scale separation, permittivity coupling and weak coupling, by using different models. These assumptions are not applicable: without time scale separation, with fast coupling and with a normal form of a saddle-node. In each case, the prediction based on a generic connectivity matrix (Supplementary Fig. 1) and the experimental score were degraded (Supplementary Fig. 4). We, however, found that by normalizing our connectivity matrix to smaller weights leads to results comparable to our model, demonstrating that weak coupling is a crucial assumption for the prediction of recruited networks.

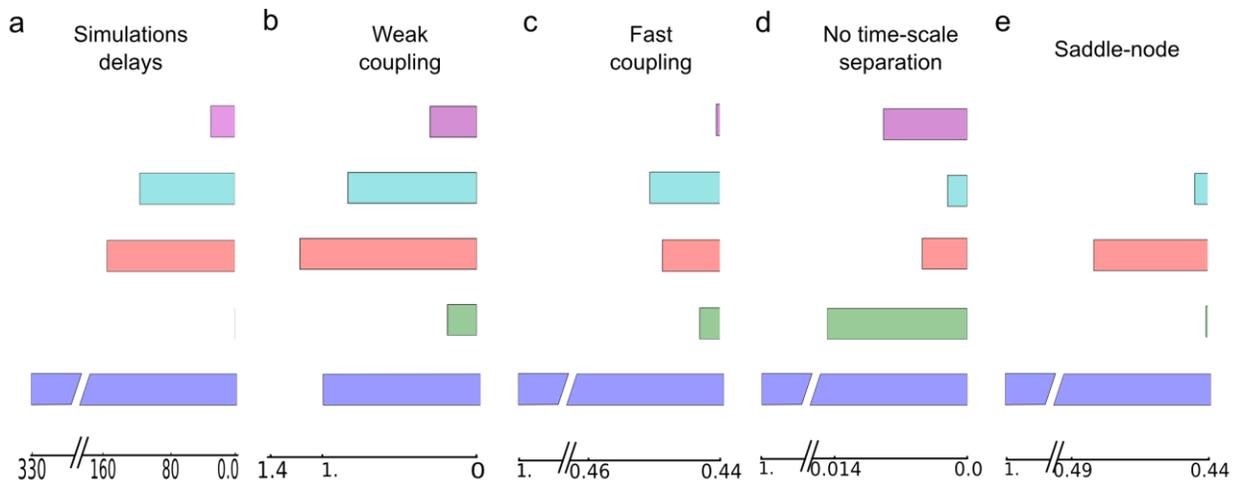

**Supplementary Figure 6: Analytical approximation and surrogate models.** Comparison of (a) the simulation delays as shown in Supplementary Fig. 1 with the absolute value of the eigenvector corresponding to the leading eigenvalue for surrogate models for (b) the analytical approximation using the weak coupling assumption, (c) the coupling on a fast time scale, (d) the model without the time-scale separation, (e) the normal form model for the generic saddle-node bifurcation. Each Epileptor node has the same color than in Supplementary Fig. 1.

## Discussion

The recruitment of distal brain regions in partial seizures is a large-scale phenomenon spanning multiple time scales. We studied the role of the large-scale connectome based on diffusion MRI, that is the structural large-scale connectivity amongst brain areas, in predicting the recruitment of distant areas by seizures originating from a focal epileptogenic network by constructing a large-scale brain network models (BNM). We demonstrated that simulations and analytical solutions approximating the BNM behavior significantly predict the propagation zone (PZ) as determined by SEEG recordings and clinical expertise. The robustness of the model predictions was examined by testing various surrogate models and connectivity matrices.

To predict the recruitment network we posited a slow permittivity difference coupling function[35], which approximates the effect of local and remote fast neuronal discharges as a perturbation of the slow permittivity variable from the local homeostatic equilibrium. Such mechanisms do not exclude additional couplings on faster time scales that could comprise spike-wave events synchronization. Synaptic and electric coupling alone fail however to explain the temporal delays up to tens of seconds long that can be observed with SEEG during network recruitments. The slow permittivity variable is supported by a variety of biophysical mechanisms that act on slow time scales such as tissue oxygenation[41], extracellular level of ions[42], and metabolism[43], which were found to vary in mouse[19], cat[44] and baboon[45] models of epileptic seizures.

Most computational models of seizure propagation often focus on small continuous spatial scales[46–48] or population of neurons[49–53]. To our best knowledge, modeling seizure propagation at a large-scale (i.e., based on diffusion MRI) was up to now only used to investigate absence seizures[21], that is, a global synchronization of the network without spatial variability. Other modeling studies focused on small networks to investigate the role of the topology and localization of the epileptogenic zone (EZ)[20]. We proposed the large-scale connectome to be a major determinant of recruitment networks, which can be investigating in large-scale brain model based on patient specific data. Diffusion MRI has revealed a quantitative decrease of regional connectivity around the EZ that is associated with a network reorganization[28–32] and cognitive impairement[54]. Histological studies provide evidence of white matter alterations in temporal lobe epilepsy[55,56]. Functional, volumetric and electrographic data suggest as well a broad reorganization of the networks in epileptic patient[57–61]. Hemispherectomy by disrupting tracts[62] have shown the interest of cutting seizure propagation pathways. Altogether, these evidences highlight the large-scale character of partial seizure propagation in the human brain. On an ad-hoc basis, clinicians routinely factor knowledge of white matter anatomy into the interpretation of SEEG data and a seizure spread but no framework exist to probe the underlying mechanisms. In this article, we used patient-specific diffusion MRI data to systematically test the relevance of the large-scale network modeling in predicting seizure recruitment networks.

Estimating the weights of the connectivity matrix by counting the streamlines among areas is controversial. The number of streamlines may not directly reflect the strength of signal transmission between two areas but simply the probability that a streamline between two regions is found by the tractography algorithm[63,64]. However, using ACT[65] and SIFT[66] methods for the tractography have been shown recently to robustly quantify structural connectivity weights and match white matter statistics estimated from post-mortem studies[67]. Additionally, we have shown that varying the weight values of the connectivity matrix have less influence on the recruited network than the topology of the connectivity matrix. As a consequence, simply those tracks that are highly likely to exist are crucial in determining the recruitment network.

The EZ and PZ estimation by the clinician is based not only on SEEG but also on prior knowledge gathered throughout the patient comprehensive presurgical evaluation[27], and can differ from direct SEEG signal power estimation. Several biomarkers are used in the daily clinical routine such as the Epileptogenicity Index[68] that based on fast discharges and delay of recruitment, or electrical stimulation to explore tissue excitability[69]. Regarding the Epileptor model, it has been shown that these two biomarkers are affected by permittivity coupling and assess the threshold level separating ictal from interictal state[19,35].

The extent of the EZ has been linked to the surgical prognosis[68,70], but is difficult to estimate. SEEG exploration uses only a limited number of electrodes (10 to 15) and therefore is sparse and can lead to poor surgical outcome if the EZ or the PZ are underestimated (Fig. 4b). By helping to predict the PZ based on the EZ, our model can support the clinician through the decision process about the localization of the epileptogenic regions in the brain, for instance by helping to discard EZ localization hypothesis that lead to spatial recruitment patterns in contradiction with SEEG recordings. This is particularly important for the choice of SEEG electrodes placement and delineating the surgical resection limits. Our model can also help to better estimate the probability of recruitment of unexplored regions, in particular subcortical regions, which are rarely considered and are linked to loss of consciousness and surgical prognosis[71]. We showed that a generic connectivity performs quite well in predicting the recruited network. It would be therefore possible to compute a catalogue of all the possible propagation patterns, accelerating the presurgical evaluation process without resorting to a dMRI scan. Finally, our results are promising but need to be validated on larger cohort of patients, in particular to examine the predictive value over surgery.

# Acknowledgement
This work was supported by the Brain Network Recovery Group through the James S. McDonnell Foundation and funding from the European Union Seventh Framework Programme (FP7-ICT BrainScales and Human Brain Project (grant no.60402)).

# References

1. Singer, W. Neuronal Synchrony: A Versatile Code for the Definition of Relations. *Neuron* **24,** 49–65 (1999).
2. Roelfsema, P. R., Engel, A. K., König, P. & Singer, W. Visuomotor integration is associated with zero time-lag synchronization among cortical areas. *Nature* **385,** 157–61 (1997).
3. Bressler, S. L. Large scale cortical networks and cognition. *Brain Res. Rev.* **20,** 288–304 (1995).
4. Engel, A. K., Konig, P., Kreiter, A. K. & Singer, W. Interhemispheric synchronization of oscillatory neuronal responses in cat visual cortex. *Science (80-. ).* **252,** 1177–1179 (1991).
5. Kelso, S. J. A. *Dynamic patterns: the self-organization of brain and behavior (Complex Adaptive Systems)*. (MIT Press, 1995).
6. Allen, E. A. *et al.* Tracking Whole-Brain Connectivity Dynamics in the Resting State. *Cereb. cortex* **24,** 663–676 (2014).
7. Hansen, E. C. a., Battaglia, D., Spiegler, A., Deco, G. & Jirsa, V. K. Functional Connectivity Dynamics: Modeling the switching behavior of the resting state. *Neuroimage* **105,** 525–535 (2014).
8. Spiegler, A., Hansen, E. C. A., Bernard, C., Mcintosh, A. R. & Jirsa, V. K. Selective activation of resting state networks following focal stimulation in a connectome- based network model of the human brain. *arXiv Prepr.*
9. Jirsa, V. K., Jantzen, K. J., Fuchs, A. & Kelso, J. a S. Spatiotemporal forward solution of the EEG and MEG using network modeling. *IEEE Trans. Med. Imaging* **21,** 493–504 (2002).
10. Jirsa, V. K., Sporns, O., Breakspear, M., Deco, G. & McIntosh, R. A. Towards the virtual brain: network modeling of the intact and the damaged brain. *Arch. Ital. Biol.* **148,** 189–205 (2010).
11. Ghosh, A., Rho, Y., McIntosh, A. R., Kötter, R. & Jirsa, V. K. Noise during rest enables the exploration of the brain's dynamic repertoire. *PLoS Comput. Biol.* **4,** e1000196 (2008).
12. Deco, G., Jirsa, V. K. & McIntosh, A. R. Emerging concepts for the dynamical organization of resting-state activity in the brain. *Nat. Rev. Neurosci.* **12,** 43–56 (2011).
13. Falcon, M. I. *et al.* Functional Mechanisms of Recovery after Chronic Stroke: Modeling with the Virtual Brain. *Eneuro* **3,** (2016).
14. Cabral, J. *et al.* Structural connectivity in schizophrenia and its impact on the dynamics of spontaneous functional networks. *Chaos* **23,** (2013).
15. Wendling, F., Bartolomei, F., Bellanger, J. J. & Chauvel, P. Epileptic fast activity can be explained by a model of impaired GABAergic dendritic inhibition. *Eur. J. Neurosci.* **15,** 1499–1508 (2002).
16. Kalitzin, S. N., Velis, D. N. & Lopes Da Silva, F. H. Stimulation-based anticipation and control of state transitions in the epileptic brain. *Epilepsy Behav.* **17,** 310–23 (2010).
17. Touboul, J., Wendling, F., Chauvel, P. & Faugeras, O. Neural mass activity, bifurcations, and epilepsy. *Neural Comput.* **23,** 3232–86 (2011).
18. Kramer, M. A. *et al.* Human seizures self-terminate across spatial scales via a critical transition. *Proc. Natl. Acad. Sci.* **109,** 21116–21121 (2012).

19. Jirsa, V. K., Stacey, W. C., Quilichini, P. P., Ivanov, A. I. & Bernard, C. On the nature of seizure dynamics. *Brain* (2014). doi:10.1093/brain/awu133

20. Terry, J. R., Benjamin, O. & Richardson, M. P. Seizure generation: the role of nodes and networks. *Epilepsia* **53,** e166–9 (2012).

21. Taylor, P. N., Goodfellow, M., Wang, Y. & Baier, G. Towards a large-scale model of patient-specific epileptic spike-wave discharges. *Biol. Cybern.* **107,** 83–94 (2013).

22. Bartolomei, F., Guye, M. & Wendling, F. Abnormal binding and disruption in large scale networks involved in human partial seizures. *EPJ Nonlinear Biomed. Phys.* 1–16 (2013).

23. Toyoda, I., Bower, M. R., Leyva, F. & Buckmaster, P. S. Early activation of ventral hippocampus and subiculum during spontaneous seizures in a rat model of temporal lobe epilepsy. *J. Neurosci.* **33,** 11100–15 (2013).

24. Talairach, J. & Bancaud, J. Lesion, 'irritative' zone and epileptogenic focus. *Confin Neurol* **27,** 4 (1966).

25. Bartolomei, F., Wendling, F., Bellanger, J. J., Régis, J. & Chauvel, P. Neural networks involving the medial temporal structures in temporal lobe epilepsy. *Clin. Neurophysiol.* **112,** 1746–60 (2001).

26. Spencer, S. S. Neural networks in human epilepsy: evidence of and implications for treatment. *Epilepsia* **43,** 219–27 (2002).

27. Bartolomei, F. *et al.* The presurgical evaluation of epilepsies. *Rev. Neurol. (Paris).* **158,** 4S55–64 (2002).

28. Ahmadi, M. E. *et al.* Side matters: diffusion tensor imaging tractography in left and right temporal lobe epilepsy. *AJNR. Am. J. Neuroradiol.* **30,** 1740–7 (2009).

29. Bernhardt, B. C., Hong, S., Bernasconi, A. & Bernasconi, N. Imaging structural and functional brain networks in temporal lobe epilepsy. *Front. Hum. Neurosci.* **7,** 624 (2013).

30. Bonilha, L. *et al.* Medial temporal lobe epilepsy is associated with neuronal fibre loss and paradoxical increase in structural connectivity of limbic structures. *J. Neurol. Neurosurg. Psychiatry* **83,** 903–909 (2012).

31. DeSalvo, M. N., Douw, L., Tanaka, N., Reinsberger, C. & Stufflebeam, S. M. Altered structural connectome in temporal lobe epilepsy. *Radiology* **270,** 842–8 (2014).

32. Besson, P. *et al.* Structural connectivity differences in left and right temporal lobe epilepsy. *Neuroimage* **100,** 135–144 (2014).

33. Sanz Leon, P. *et al.* The Virtual Brain: a simulator of primate brain network dynamics. *Front. Neuroinform.* **7,** 1–23 (2013).

34. Sanz-Leon, P., Knock, S. A., Spiegler, A. & Jirsa, V. K. Mathematical framework for large-scale brain network modelling in The Virtual Brain. *NeuroImage.* **111,** 385–430 (2015).

35. Proix, T., Bartolomei, F., Chauvel, P., Bernard, C. & Jirsa, V. K. Permittivity Coupling across Brain Regions Determines Seizure Recruitment in Partial Epilepsy. *J. Neurosci.* **34,** 15009–15021 (2014).

36. Litt, B., Esteller, R., Echauz, J. & D'Alessandro, M. Epileptic seizures may begin hours in advance of clinical onset: a report of five patients. *Neuron* **30,** 51–64 (2001).

37. Engel, J. & Engel. *Surgical Treatment of the Epilepsies*. (Lippincott Williams & Wilkins, 1993).

38. Watts, D. J. & Strogatz, S. H. Collective dynamics of 'small-world' networks. *Nature* **393,** 440–2

(1998).

39. Hagmann, P. *et al.* Mapping human whole-brain structural networks with diffusion MRI. *PLoS One* **2,** e597 (2007).

40. Hagmann, P. *et al.* Mapping the structural core of human cerebral cortex. *PLoS Biol.* **6,** e159 (2008).

41. Suh, M., Hongtao, M., Mingrui, Z., Sharif, S. & Schwartz, T. H. Neurovascular coupling and oximetry during epileptic events. *Mol. Neurobiol.* **33,** 181–197 (2006).

42. Heinemann, U., Konnerth, A., Pumain, R. & Wadman, W. J. Extracellular calcium and potassium concentration changes in chronic epileptic brain tissue. *Adv. Neurol.* **44,** 641–61 (1986).

43. Zhao, M. *et al.* Preictal and ictal neurovascular and metabolic coupling surrounding a seizure focus. *J. Neurosci.* **31,** 13292–300 (2011).

44. Moody, W. J., Futamachi, K. J. & Prince, D. A. Extracellular potassium activity during epileptogenesis. *Exp. Neurol.* **42,** 248–263 (1974).

45. Pumain, R., Menini, C., Heinemann, U., Louvel, J. & Silva-Barrat, C. Chemical synaptic transmission is not necessary for epileptic seizures to persist in the baboon Papio papio. *Exp. Neurol.* **89,** 250–8 (1985).

46. Hall, D. & Kuhlmann, L. Mechanisms of seizure propagation in 2-dimensional centre-surround recurrent networks. *PLoS One* **8,** e71369 (2013).

47. Ursino, M. & La Cara, G.-E. Travelling waves and EEG patterns during epileptic seizure: analysis with an integrate-and-fire neural network. *J. Theor. Biol.* **242,** 171–87 (2006).

48. Kim, J. W., Roberts, J. A. & Robinson, P. A. Dynamics of epileptic seizures: evolution, spreading, and suppression. *J. Theor. Biol.* **257,** 527–532 (2009).

49. Miles, R., Traub, R. D. & Wong, R. K. Spread of synchronous firing in longitudinal slices from the CA3 region of the hippocampus. *J Neurophysiol* **60,** 1481–1496 (1988).

50. Golomb, D. & Amitai, Y. Propagating neuronal discharges in neocortical slices: computational and experimental study. *J. Neurophysiol.* **78,** 1199–211 (1997).

51. Compte, A., Sanchez-Vives, M. V, McCormick, D. a & Wang, X.-J. Cellular and network mechanisms of slow oscillatory activity (<1 Hz) and wave propagations in a cortical network model. *J. Neurophysiol.* **89,** 2707–25 (2003).

52. Bazhenov, M., Timofeev, I., Fröhlich, F. & Sejnowski, T. J. Cellular and network mechanisms of electrographic seizures. *Drug Discov. Today. Dis. Models* **5,** 45–57 (2008).

53. Fröhlich, F., Bazhenov, M., Timofeev, I. & Sejnowski, T. J. Maintenance and termination of neocortical oscillations by dynamic modulation of intrinsic and synaptic excitability. *Thalamus Relat. Syst.* **3,** 147 (2007).

54. Leyden, K. M. *et al.* What does diffusion tensor imaging ( DTI ) tell us about cognitive networks in temporal lobe epilepsy ? *Quant. Imaging Med. Surg.* **5,** 247–263 (2015).

55. Thom, M., Sisodiya, S., Harkness, W. & Scaravilli, F. Microdysgenesis in temporal lobe epilepsy. A quantitative and immunohistochemical study of white matter neurones. *Brain* **124,** 2299–2309 (2001).

56. Blanc, F. *et al.* Investigation of widespread neocortical pathology associated with hippocampal

sclerosis in epilepsy: A postmortem study. *Epilepsia* **52,** 10–21 (2011).

57. Lieb, J., Hoque, K., Skomer, C. & Song, X. Inter-hemispheric propagation of human mesial temporal lobe seizures: a coherence/phase analysis. *Electroencephalogr. Clin. Neurophysiol.* **67,** 101–119 (1987).

58. Lieb, J. P., Dasheiff, R. M. & Engel, J. Role of the frontal lobes in the propagation of mesial temporal lobe seizures. *Epilepsia* **32,** 822–37 (1991).

59. Cassidy, R. M. & Gale, K. Mediodorsal thalamus plays a critical role in the development of limbic motor seizures. *J. Neurosci.* **18,** 9002–9 (1998).

60. Rosenberg, D. S. *et al.* Involvement of medial pulvinar thalamic nucleus in human temporal lobe seizures. *Epilepsia* **47,** 98–107 (2006).

61. Bettus, G. *et al.* Decreased basal fMRI functional connectivity in epileptogenic networks and contralateral compensatory mechanisms. *Hum. Brain Mapp.* **30,** 1580–91 (2009).

62. Beier, A. D. & Rutka, J. T. Hemispherectomy: historical review and recent technical advances. *Neurosurg. Focus* **34,** E11 (2013).

63. Jbabdi, S. & Johansen-Berg, H. Tractography: Where Do We Go from Here? *Brain Connect.* **1,** 169–183 (2011).

64. Jones, D. K., Knösche, T. R. & Turner, R. White matter integrity, fiber count, and other fallacies: The do's and don'ts of diffusion MRI. *Neuroimage* **73,** 239–254 (2013).

65. Smith, R. E., Tournier, J.-D., Calamante, F. & Connelly, A. Anatomically-constrained tractography: improved diffusion MRI streamlines tractography through effective use of anatomical information. *Neuroimage* **62,** 1924–38 (2012).

66. Smith, R. E., Tournier, J.-D., Calamante, F. & Connelly, A. SIFT: Spherical-deconvolution informed filtering of tractograms. *Neuroimage* **67,** 298–312 (2013).

67. Smith, R. E., Tournier, J.-D., Calamante, F. & Connelly, A. The effects of SIFT on the reproducibility and biological accuracy of the structural connectome Article. *Neuroimage* **104,** 253–265 (2015).

68. Bartolomei, F. *et al.* From mesial temporal lobe to temporoperisylvian seizures: a quantified study of temporal lobe seizure networks. *Epilepsia* **51,** 2147–58 (2010).

69. Valentín, A. *et al.* Single pulse electrical stimulation for identification of structural abnormalities and prediction of seizure outcome after epilepsy surgery: A prospective study. *Lancet Neurol.* **4,** 718–726 (2005).

70. Aubert, S. *et al.* Local and remote epileptogenicity in focal cortical dysplasias and neurodevelopmental tumours. *Brain* **132,** 3072–86 (2009).

71. Arthuis, M. *et al.* Impaired consciousness during temporal lobe seizures is related to increased long-distance cortical-subcortical synchronization. *Brain* **132,** 2091–101 (2009).

72. Proix, T. *et al.* How do parcellation size and short-range connectivity affect large-scale brain network models? *(submitted)* (2016).

73. Fischl, B. FreeSurfer. *Neuroimage* **62,** 774–81 (2012).

74. Jenkinson, M., Beckmann, C. F., Behrens, T. E. J., Woolrich, M. W. & Smith, S. M. Fsl. *Neuroimage* **62,** 782–90 (2012).

75. Tournier, J.-D. MRtrix package, Brain Research Institute, Melbourne, Australia,

https://github.com/jdtournier/mrtrix3. Available at: https://github.com/jdtournier/mrtrix3.

76. Fuhrmann, S., Ackermann, J., Kalbe, T. & Goesele, M. Direct Resampling for Isotropic Surface Remeshing. in *Vision, Modeling and Visualization* (2010).
77. Desikan, R. S. *et al.* An automated labeling system for subdividing the human cerebral cortex on MRI scans into gyral based regions of interest. *Neuroimage* **31,** 968–80 (2006).
78. Zalesky, A. *et al.* Whole-brain anatomical networks: does the choice of nodes matter? *Neuroimage* **50,** 970–83 (2010).
79. Tournier, J.-D., Calamante, F. & Connelly, A. Robust determination of the fibre orientation distribution in diffusion MRI: non-negativity constrained super-resolved spherical deconvolution. *Neuroimage* **35,** 1459–72 (2007).
80. Haken, H. *Information and Self-Organization*. (Springer Berlin Heidelberg, 2000).
81. El Houssaini, K., Ivanov, A. I., Bernard, C. & Jirsa, V. K. Seizures, refractory status epilepticus, and depolarization block as endogenous brain activities. *Phys. Rev. E* **91,** 2–6 (2015).
82. Haken, H. *Advanced Synergetics: Instability Hierarchies of Self-Organizing Systems and Devices*. (Springer, 1987).

# Online Methods

## Patient selection and data acquisition

We selected 15 drug-resistant patients (6 males, mean age 33.4, range 22-56) with different types of partial epilepsy accounting for different epileptogenic zone localizations. All patients underwent a presurgical evaluation (Supplementary Table 1). The first phase in the evaluation of each patient is not invasive and comprises the patient clinical record, neurological examinations, positron emission tomography (PET), and electroencephalography (EEG) along with video monitoring. T1 weighted anatomical images (MPRAGE sequence, TR=1900 ms, TE=2.19 ms, 1.0 x 1.0 x 1.0 mm, 208 slices) and diffusion MRI images (DTI-MR sequence, angular gradient set of 64 directions, TR=10.7 s, TE=95 ms, 2.0 x 2.0 x2.0 mm, 70 slices, b weighting of 1000 s/mm$^2$) were also acquired on a Siemens Magnetom Verio 3T MR-scanner. From the gathered data clinicians conclude potential epileptogenic zones (EZ). Further elaboration on the EZ are done in the second phase, which is invasive and comprises the placement of stereotactic EEG (SEEG) electrodes in or close to the suspected regions. These electrodes have 10 to 15 contacts that are 1.5 mm apart. Each contact is 2 mm of length and 0.8 mm in diameter. The SEEG was recorded by a 128 channel Deltamed$^{TM}$ system using a 256 Hz sampling rate. The SEEG recordings were band-pass filtered between 0.16 and 97 Hz by a hardware filter. All the chosen patients showed seizures in the SEEG starting in one or several localized areas, that is, the EZ before recruiting distant regions, that is, the propagation zone (PZ). The position of the electrodes was pinned down by performing a computerized tomography (CT) scan or a MRI after implanting the electrodes.

Additionally, five healthy controls signed an informed consent form according to the rules of the local ethics committee (Comité de Protection des Personnes (CPP) Marseille 2) and underwent the same MRI protocol.

## Data processing

To import structural and diffusion MRI data in The Virtual Brain the data was processed using SCRIPTS[72]. This processing pipeline makes use of various tools such as FreeSurfer[73], FSL[74], MRtrix3[75] and Remesher[76] to reconstruct the individual cortical surface and large-scale connectivity. The surface was reconstructed using 20,000 vertices. Cortical and volumetric parcellations were performed using the Desikan-Killiany atlas with 70 cortical regions and 17 subcortical regions[77]. Two additional parcellations were obtained by subdividing each cortical regions of the Desikan-Killiany atlas in two and four to obtain 157 and 297 regions, respectively[78]. The diffusion data was corrected for eddy-currents and head motions using *eddy-correct* FSL functions. Fiber orientation estimation was performed with Constrained Spherical Deconvolution[79], and improved with Anatomically Constrained Tractography[65]. Tractography was performed using $2.5 \cdot 10^6$ fibers and were corrected using Spherical-Deconvolution Informed Filtering of Tractograms[66]. The connectivity matrix was obtained by summing track counts over each region of the parcellation, and normalized so that the maximum value of the connectivity matrix was one.

The CT or MRI scan performed after electrode placements were aligned with the structural MRI recorded before the surgery using the FLIRT function of FSL, with 6 degrees of freedom and a Mutual Information cost function. Each contact surface was reconstructed and assigned to the region of the corresponding parcellations containing the most of the contact volume.

## Definition of the Propagation Zone

The propagation was defined by two different methods. The first method is the subjective evaluation of clinicians based on the different measurement modalities (EEG and SEEG) gathered throughout the two-step procedure (non-invasive and invasive). The second method is objective in the sense that it simply based on the SEEG. For each patient, all seizures were isolated in the SEEG time series. The bipolar SEEG was considered (between pairs of electrode contacts) and filtered between 1-50 Hz using a Butterworth band-pass filter. A contact was considered to be in the PZ if its signal energy was responsible for at least 30% of the maximum signal energy over the contacts, and was not in the EZ. The corresponding region was then assigned to the PZ

## Comparing the estimates of the PZ

Two different scores were computed to compare predicted PZ with the estimated PZ as describe above. The binary score $S_1$ simply counts the accordance of each region found in the simulated PZ (ensemble $PZS$) and the predicted PZ (ensemble $PZP$):

$$S_1 = \sum_{i=1}^{N_{PZS}} \begin{cases} 1 & \text{if } R_i \in PZP \\ 0 & \text{if } R_i \notin PZP \end{cases}$$

The distance score $S_2$ quantitatively estimates the L1-norm between the normalized probability of the predicted propagation and the strength of the SEEG signal power (for clinician prediction, the strength was set to 1):

$$S_2 = \sum_{i=1}^{N_{PZS}} (1 - |x_{PZS}^i - x_{PZP}^i|)$$

## Chance level to get areas in the PZ

The chance level was the probability of obtaining $k$ regions in the PZ by drawing randomly $n$ regions from the set of all regions in the parcellations, multiplied by the score obtained for $k$ regions, summed over all possibilities for $k$ [80]. This can be mathematically written as:

$$\sum_{k=1}^{m} \text{score}(k) \cdot P(X = k) = \sum_{k=1}^{m} \frac{k}{m} \cdot \frac{\binom{m}{k} \cdot \binom{N-m}{n-k}}{\binom{N}{m}},$$

with $m \in \mathbb{Z}$ the number of regions that are in the clinical estimation of the PZ, $n$ the number of regions drawn randomly from the parcellations, $N$ the total number of regions in the parcellation.

## Modeling

### Brain Network Model

Based on recordings of epileptic seizures in different species, Jirsa et al[19] identified the dominant bifurcations involved at seizure onset and offset amid the theoretically 16 possible classes of bifurcation pairs for bursting activity predicted by dynamical system theory. This consideration resulted in a phenomenological model, called the Epileptor. The activity in this model is autonomously switching between interictal and ictal states because of a slow permittivity variable that is supposed to be related to tissue oxygenation[41], metabolism[43], and extracellular levels of ions[42]. The Epileptor is mathematically a set of two coupled oscillators linked together by the slow permittivity variable $z$. The two oscillators account for the fast discharges (variables $x_1$ and $y_1$) and spike and wave events (variables $x_2$ and $y_2$) observed in electrogaphic seizure recordings. The Epileptor model was also shown to reproduce refractory status epilepticus and depolarization block[81]. Using time-scale separation as well as evidences from SEEG recordings, Proix et al.[35] suggested to consider seizure recruitment among brain regions on a slow time scale. Such a coupling function can be expressed as a linear difference coupling term, subsuming first order deviations from the homeostatic equilibrium of the slow permittivity variable. We used the same approach here in the general case of $N$ coupled Epileptors, which reads for each Epileptor $i$:

$$\dot{x}_{1,i} = y_{1,i} - f_1(x_{1,i}, x_{2,i}) - z_i + I_1$$
$$\dot{y}_{1,i} = 1 - 5x_{1,i}^2 - y_{1,i}$$
$$\dot{z}_i = \frac{1}{\tau_0}\left(4(x_{1,i} - x_{0,i}) - z_i - \sum_{j=1}^{N} K_{ij} \cdot (x_{1,j} - x_{1,i})\right)$$
$$\dot{x}_{2,i} = -y_{2,i} + x_{2,i} - x_{2,i}^3 + I_2 + 0.002g(x_{1,i}) - 0.3(z_i - 3.5)$$
$$\dot{y}_{2,i} = \frac{1}{\tau_2}(-y_{2,i} + f_2(x_{2,i}))$$

where

$$f_1(x_{1,i}, x_{2,i}) = \begin{cases} x_{1,i}^3 - 3x_{1,i}^2 & \text{if } x_{1,i} < 0 \\ (x_{2,i} - 0.6(z_i - 4)^2)x_{1,i} & \text{if } x_{1,i} \geq 0 \end{cases}$$

$$f_2(x_{2,i}) = \begin{cases} 0 & \text{if } x_{2,i} < -0.25 \\ 6(x_{2,i} + 0.25) & \text{if } x_{2,i} \geq -0.25 \end{cases}$$

$$g(x_{1,i}) = \int_{t_0}^{t} e^{-\gamma(t-\tau)} x_{1,i}(\tau) d\tau$$

and $\tau_0 = 2857; \tau_2 = 10; I_1 = 3.1; I_2 = 0.45; \gamma = 0.01$. The degree of excitability of each Epileptor is represented by the value $x_{0,i}$ that we varied in this study. To simplify the interpretation, we define $\Delta x_{0,i} = x_{0,i} - x_0^C$ with $x_0^C = -2.1$ the critical value of excitability. If $\Delta x_0 > 0$, a brain region is epileptogenic and seizures are triggered autonomously. Otherwise, $\Delta x_0 < 0$ and regions are in a healthy equilibrium state.

### Numerical simulations
The simulations were performed with The Virtual Brain[33] using a Heun integration scheme (time step: 0.04 ms). A zero mean white Gaussian noise with a variance of $0.0025$ was added to the variables to make time series resemble more closely the SEEG recorded seizures. $x_2$ and $y_2$. 256 time steps correspond to 1s of physical time for realistic seizure durations. In Fig. 2, simulated signals are filtered with an order 5 bandpass Butterworth filter (0.16 Hz-97 Hz) to reproduce hardware filters used in SEEG data acquisition.

## Mathematical analysis

### Two-dimensional reduction of the Epileptor

Taking advantage of time scale separation and focusing on the slower time scale, the five dimensional Epileptor reduces to[35]:

$$\begin{cases} \dot{x}_{1,i} = -x_{1,i}^3 - 2x_{1,i}^2 + 1 - z_i + I_{1,i} \\ \dot{z}_i = \frac{1}{\tau_0}\left(4(x_{1,i} - x_{0,i}) - z_i - \sum_{j=1}^{N} K_{ij}(x_{1,j} - x_{1,i})\right) \end{cases}$$

for each Epileptor $i$ with $\tau_0 = 2857$ and $I_{1,i} = 3.1$.

### Reduction of the Epileptor to the slow manifold

In detail, we applied the averaging methods[82] to project the system on the slow manifold by setting $\dot{x}_i = 0$, which gives $x_i = F(z_i)$. This approximation holds as long as $\tau_0 \gg 1$, otherwise the approximation breaks down as shown for a generic connectivity matrix in Supplementary Fig. 6d (no time-scale separation), and for a patient connectivity matrix in Supplementary Fig. 4(time-scale). We computed $F(z_i)$ explicitly: first we approximated the third order polynomial by a second order polynomial using a second order Taylor expansion in $x = -4/3$, giving $\dot{x}_i \approx 2x_i^2 + 16/3 x_i + 4.1 + 64/27 - z_i$. Setting $\dot{x}_i = 0$, we obtained

$$F(z_i) = 1/4(-16/3 - \sqrt{8z_i - 629.6/27})$$

The 2$N$ dimensional system then becomes 1$N$ dimensional, and read as follows:

$$\dot{z} = \frac{1}{\tau_0}(-4x_0 + 4F(z_i) - z - \sum_{j=1}^{N} K_{ij}(F(z_j) - F(z_i)))$$

### Linear stability analysis

We estimated the PZ by identifying the dominating sub-networks involved in the transition toward the seizure state via a linear stability analysis. We chose the excitability values of the EZ, of the other nodes, and the coupling strength such that the nodes in the EZ are close to the system separatrix. We computed the fixed point solution $\bar{z}_i, \forall i \in [1, N]$ of the system by setting $\dot{z}_i = 0$. Then we computed numerically the dominant eigenvalues and of the Jacobian:

$$J = \frac{1}{\tau_0} \begin{bmatrix} d(1) & e(1,2) & \cdots & e(1,N) \\ e(2,1) & d(2) & \cdots & e(2,N) \\ \vdots & \vdots & \ddots & \vdots \\ e(N,1) & e(N,2) & \cdots & d(N) \end{bmatrix}$$

With $d(i) = -1 + F'(\bar{z}_i) \cdot (4 + \sum_{j=1}^{N} K_{ij})$, $e(i,j) = -K_{ij} \cdot F'(\bar{z}_j)$, and $F'(z) = -1/\sqrt{8z - 629.6/27}$. The probability of a node to be recruited is then given by the absolute value of the sum of the k first eigenvectors, where k is the number of nodes in the ZE.

### Weak Coupling

We further reduced the system to obtain an analytical expression of the leading eigenvectors. Let us consider the generic model:

$$\dot{Z}_i = G(Z_i) + X_{0,i} + \varepsilon \sum_{j=1}^{N} K_{ij} H(Z_i, Z_j)$$

with $\varepsilon \ll 1$. To find the fixed point solution, we note that the coupling term $\varepsilon \sum_{j=1}^{N} K_{ij} H(Z_j)$ is weak in front of the other terms. This is the assumption of weak coupling. In particular this holds for the difference coupling term $H(Z_i, Z_j) = K_{ij}(Z_j - Z_i)$. The fixed point solution is given in a first order approximation by setting $\dot{Z}_i = 0$, which gives $\bar{Z}_i = G^{-1}(-X_{0,i})$, that is, the fixed point of the uncoupled system. We search for the leading eigenvector, assuming the first region is the EZ (in case of $k$ regions in the EZ, the same reasoning is valid, but we have to consider $k$ eigenvectors instead). We assume the leading eigenvector $N = (v_1, \ldots, v_n)$ will be small at the first order for all components except for the epileptogenic region, whose coordinate is arbitrarily set to $v_1 = 1$. Writing the Jacobian and the system for the leading eigenvector read as follows:

$$(J - \Lambda I) \cdot V = 0 \quad (2)$$

With $I$ the identity matrix,

$$J = \begin{bmatrix} d(1) & e(1,2) & \cdots & e(1,N) \\ e(2,1) & d(2) & \cdots & e(2,N) \\ \vdots & \vdots & \ddots & \vdots \\ e(N,1) & e(N,2) & \cdots & d(N) \end{bmatrix}$$

$$d(i) = G'(G^{-1}(-X_{0,i})) - \lambda_1 + \varepsilon \, \partial H/\partial Z_i \, (G^{-1}(-X_{0,i}) \sum_{j=1}^{N} K_{ij})$$

$$e(i,j) = \varepsilon \sum_{j=1}^{N} K_{ij} \, \partial H/\partial Z_j \, (G^{-1}(-X_{0,j}))$$

In the first equation of the system, each term is of order 2, except for the term in $v_1$ and since $v_1 = 1$, we have $\lambda_1 = d(1)$. We can then calculate iteratively all the other terms:

$$v_j = -\frac{e(j,1)}{d(j) - \lambda_1}$$

Therefore all components $v_i$ (other than $v_1$) are small at first order. For our BNM of the Epileptors, $\partial H / \partial Z_j = 1$, $G(Z_i) = 4F(Z_i) - Z_i$, and $G'(Z_i) = 4F'(Z_i) - 1 = \frac{-1}{\sqrt{8Z_i + 629.6/27}} - 1$. We checked numerically that this analytical expression is valid by computing for the same connectivity matrix the full system and the system assuming weak coupling (Supplementary Fig. 6a-b). We also checked that the components $v_i$ (other than $v_1$) are of second order. A consequence for real connectivity matrices is that, since the distribution of weights is log-normal, some connections have a high weight for each node, and will systematically have an important value in the eigenvector: a region well connected is a region well recruited. Note that if $X_{0,i} = c, \forall i \notin EZ$, with $c$ a constant, as it is our case for the real patient data, then the PZ is directly determined by the out-strength of the EZ.

## Surrogate models

### Fast coupling

The coupling function was operating on the fast time scale, with opposite sign to keep the coupling function excitatory:

$$\begin{cases} \dot{x}_{1,i} = -x_{1,i}^3 - 2x_{1,i}^2 + 1 - z_i + I_{1,i} + \sum_{j=1}^{N} K_{ij}\left(x_{1,j} - x_{1,i}\right) \\ \dot{z}_i = \frac{1}{\tau_0}\left(4\left(x_{1,i} - x_{0,i}\right) - z_i\right) \end{cases}$$

**Time-scale separation**

The time-scale separation was suppressed by setting $\tau_0 = 1$ in system .

*Generic saddle-node bifurcation*

We also used a normal form of a saddle-node bifurcation as a generic model.

$$\dot{x}_i = x_i^2 + x_{0,i} + 2.1 + \sum_{j=1}^{N} K_{ij} x_{1,j}$$

## Surrogate connectivities

### Shuffled connectivity

Cortical regions of each hemisphere were shuffled separately. A pair of regions was chosen randomly, and the intra-hemispheric connections of this pair were switched by exchanging columns and rows (diagonal terms of the connectivity matrix were set to 0). This operation was repeated $N$ times, where $N$ was the number of regions in the hemisphere, to get one shuffle connectivity matrix. Subcortical connectivity was left intact to lessen the changes in the shuffled connectivity matrix.

### Changing the weights

Each non-zero weight $k_{ij}$ of the connectivity was summed to a draw of a uniform random distribution, whose values were taken between $-\varepsilon \cdot k_{ij}$ and $+\varepsilon \cdot k_{ij}$, with $\varepsilon$ a chosen percentage.

### Log of the connectivity matrix

We simply redefined $\tilde{k}_{ij} = \log(k_{ij} + 1)$ as the log of the connectivity matrix.